\documentclass[10pt,preprintnumbers,amsmath,amssymb,floatfix,twocolumn,prb,superscriptaddress]{revtex4}
\usepackage{mathrsfs}
\usepackage{bm}
\usepackage{amsmath}
\usepackage[dvips]{graphics,color}
\usepackage{graphicx}
\usepackage{epstopdf}
\usepackage{dcolumn}
\usepackage{bm}
\usepackage{longtable}
\usepackage{graphics}
\usepackage{amssymb}
\usepackage{xspace}
\usepackage{epsfig}
\usepackage{subfigure}

\begin{document}
\title{Electron-electron interactions in graphene bilayers}
\author{Fan Zhang}
\email{zhangfan@physics.utexas.edu} \affiliation{Department of
Physics, University of Texas at Austin, Austin TX 78712, USA}
\author{Hongki Min}
\affiliation{Department of Physics, University of Texas at Austin,
Austin TX 78712, USA}
\author{Marco Polini}
\affiliation{NEST-CNR-INFM and Scuola Normale Superiore, I-56126
Pisa, Italy}
\author{A.H. MacDonald}
\affiliation{Department of Physics, University of Texas at Austin,
Austin TX 78712, USA}

\maketitle {\bf Electrons most often organize into Fermi-liquid
states in which electron-electron interactions play an inessential
role.  A well known exception is the case of one-dimensional (1D)
electron systems (1DES).  In 1D the electron Fermi-surface consists
of points, and divergences associated with low-energy particle-hole
excitations abound when electron-electron interactions are described
perturbatively. In higher space dimensions, the corresponding
divergences occur only when Fermi lines or surfaces satisfy
idealized nesting conditions. In this article we discuss
electron-electron interactions in 2D graphene bilayer systems which
behave in many ways as if they were one-dimensional, because they
have Fermi points instead of Fermi lines and because their
particle-hole energies have a quadratic dispersion which compensates
for the difference between 1D and 2D phase space. We conclude, on
the basis of a perturbative RG calculation similar to that commonly
employed in 1D systems, that interactions in neutral graphene
bilayers can drive the system into a strong-coupling broken symmetry
state with layer-pseudospin ferromagnetism and an energy gap. }
\\\\

\begin{figure}[htbp]
\centering \scalebox{0.6}
{\includegraphics*[1.9in,5.6in][7.4in,9.25in]{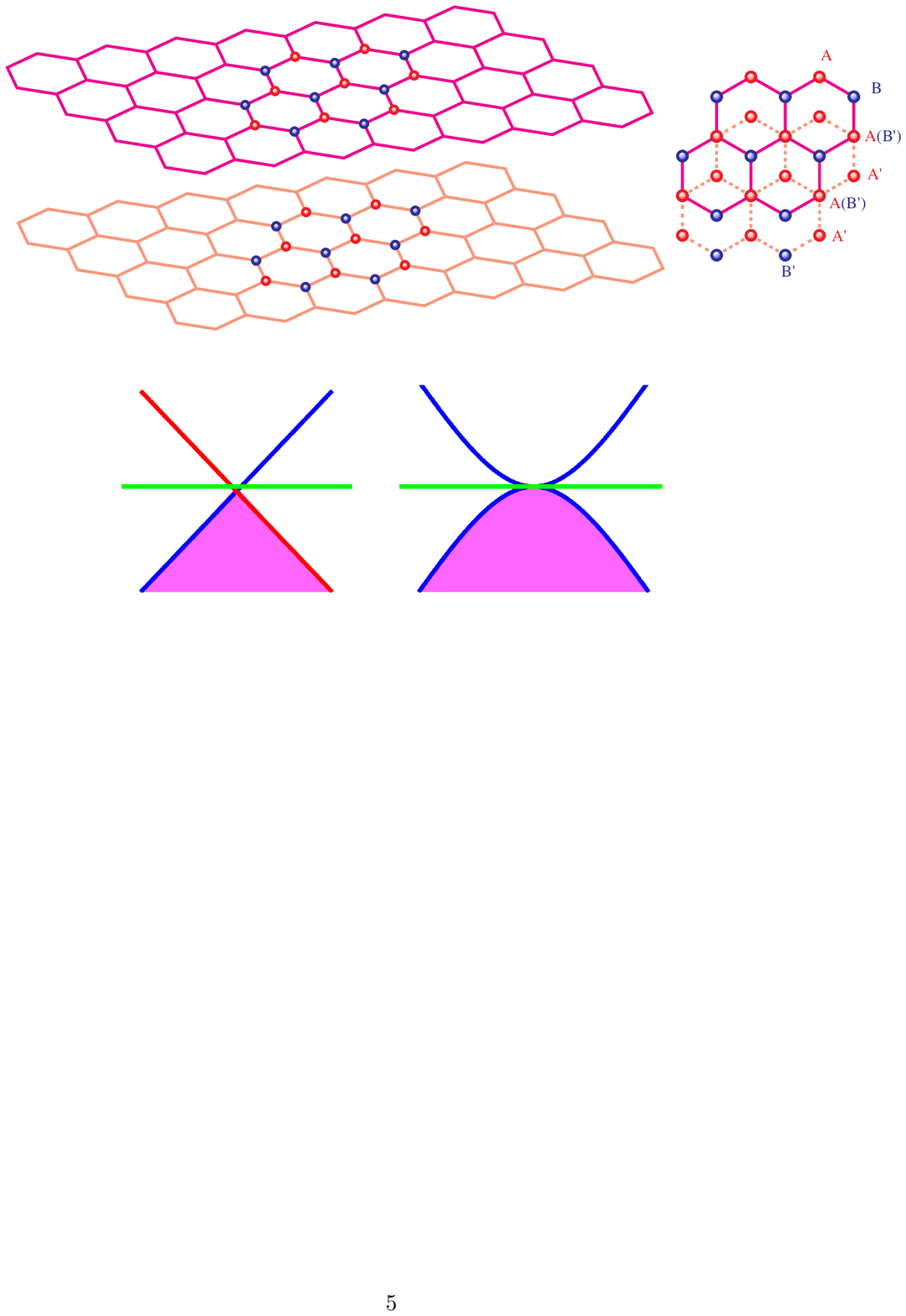}}
\caption{\label{fig:one} {{\bf Bilayer graphene lattice structure
and electronic structure.}  The massive chiral fermion model
describes the low-energy sites in a AB-stacked graphene bilayer,
those atom sites (top layer $B$ sites and bottom layer $A'$ sites)
which do not have a neighbor in the opposite layer. The conduction
and valence bands touch at the Brillouin-zone corner wavevectors,
taken as zero-momentum in continuum model theories, and separate
quadratically with increasing wavevector. In a 1DES left and right
going electrons cross the Fermi energy at a single point. In this
figure the momentum of right-going (left-going) electrons is plotted
relative to $+k_F$ ($-k_F$) where $k_F$ is the Fermi wavevector.}}
\end{figure}

Recent progress in the isolation of nearly perfect single and
multilayer graphene
sheets\cite{graphene_reviews_1,graphene_reviews_2,graphene_reviews_3,graphene_reviews_4}
has opened up a new topic in two-dimensional electron systems (2DES)
physics.  There is to date little unambiguous experimental evidence
that electron-electron interactions play an essential role in the
graphene family of 2DES's.  However, as pointed out by Min {\em et
al.}\cite{min_prb_2008} graphene bilayers near neutrality should be
particularly susceptible to interaction effects because of their
peculiar massive-chiral\cite{mccann} band Hamiltonian, which has an
energy-splitting between valence and conduction bands that vanishes
at ${\bm k} = {\bm 0}$ and grows quadratically with $k=|{\bm k}|$:
\begin{equation}
\label{eq:band} {\cal H}_{B}=-\sum_{{\bm k}\sigma'\sigma}
\frac{\hbar^2 k^2}{2m^*}c^{\dagger}_{{\bm k}\sigma'}\big[\cos(J
\phi_{\bm k})\tau^{x}_{\sigma'\sigma}+\sin(J\phi_{\bm k})
\tau^{y}_{\sigma'\sigma}\big]c_{{\bm k}\sigma}.
\end{equation}
In Eq.~(\ref{eq:band}) the $\tau^{i}$s are Pauli matrices and the
Greek labels refer to the two bilayer graphene sublattice sites, one
in each layer, which do not have a neighbor in the opposite graphene
layer.  (See Fig.~\ref{fig:one}.)  The other two sublattice site
energies are repelled from the Fermi level by interlayer hopping and
irrelevant at low energies.  It is frequently useful to view quantum
two-level layer degree of freedom as a pseudospin.  The $J=2$
pseudospin chirality of bilayer graphene contrasts with the $J=1$
chirality\cite{graphene_reviews_1, chirality_correlations} of
single-layer graphene and is a consequence of the two-step process
in which electrons hop between low-energy sites via the high-energy
sites. The massive-chiral band-structure model applies at energies
smaller than the interlayer hopping scale\cite{graphene_reviews_4}
$\gamma_1 \sim 0.3~{\rm eV}$ but larger than the trigonal-warping
scale\cite{graphene_reviews_4} $\gamma_3 \gamma_1/\gamma_0 \sim
0.03~{\rm eV}$ below which direct hopping between low-energy sites
plays an essential role.  The body of this paper concerns the role
of interactions in the massive-chiral model; we return at the end to
explain the important role played by trigonal warping.

Similarities and differences between graphene bilayers and 1DES are
most easily explained by temporarily neglecting the spin, and in the
case of graphene also the additional valley degree of freedom. As
illustrated in Fig.~\ref{fig:one} in both cases the Fermi sea is
point-like and there is a gap between occupied and empty
free-particle states which grows with wavevector, linearly in the
1DES case. These circumstances are known to support a mean-field
broken symmetry state in which phase coherence is established
between conduction and valence band states for arbitrarily weak
repulsive interactions.  In the case of 1DES, the broken symmetry
state corresponds physically to a charge density-wave (CDW) state,
while in the case of bilayer graphene\cite{min_prb_2008} it
corresponds to state in which charge is spontaneously transferred
between layers.  This mean-field theory prediction is famously
incorrect in the 1DES case, and the origin of the failure can be
elegantly identified\cite{Giamarchi,RGShankar} using a perturbative
renormalization group (PRG) approach.  We show below that when
applied to bilayer graphene, the same considerations lead to a
different conclusion.
\renewcommand{\thefootnote}{\fnsymbol{footnote}}

The reliability of the mean-field theory
prediction\cite{min_prb_2008} of a weak-interaction instability in
bilayer graphene can be systematically assessed using
PRG\cite{RGShankar}. We outline the main steps in the application of
this analysis to bilayer graphene in the main text, pointing out
essential differences compared to the 1DES case. Details are
provided in the supplementary material. We assume short-range
interactions\footnotemark[1]\footnotetext[1]{We replace the bare
Coulomb interactions by short-range momentum-independent
interactions\cite{RGShankar} by evaluating them at typical momentum
transfers at the model's high-energy limit. We believe that this
approximation is not serious because of screening.} between
electrons in the same ($S$) and different ($D$) layers.

\begin{figure}[htbp]
\centering {\scalebox{0.5}
{\includegraphics*[1.8in,5.35in][8.2in,9.4in]{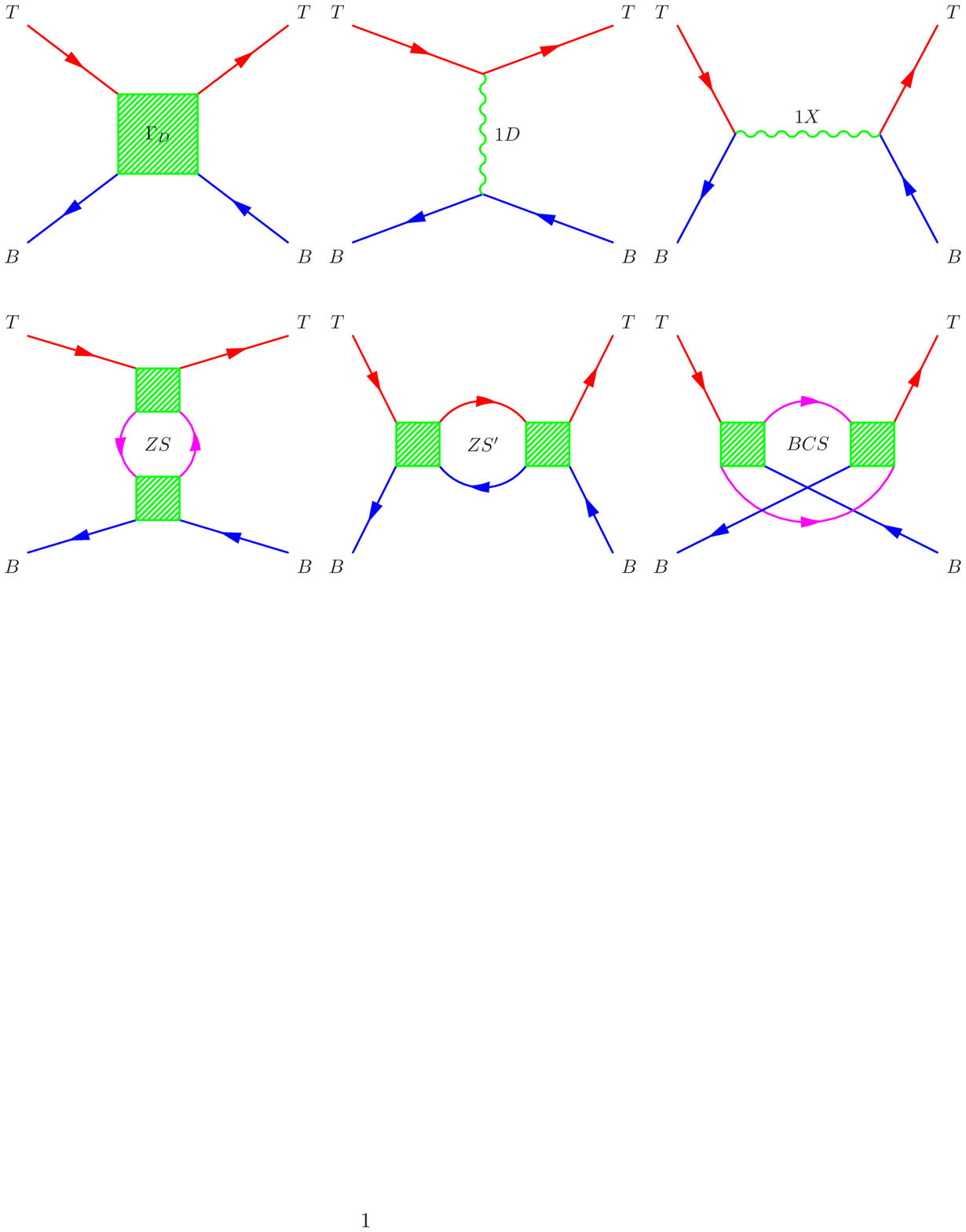}}}
\caption{\label{fig:three} {{\bf Four-point scattering function for
bilayer graphene.} The 1st diagram is the renormalized interaction
while the 2nd and 3rd ones are the direct and exchange bare
interactions, respectively. The following diagrams are the one-loop
diagrams labelled ZS, ZS' and BCS. The external and internal Green's
function labels refer to layer in the case of graphene and to
chirality in the case of 1DES. }}
\end{figure}

The PRG analysis centers on the four point scattering function
defined in terms of Feynman diagrams in Fig.~\ref{fig:three}. Since
the Pauli exclusion principle implies that (in the spinless
valleyless case) no pair of electrons can share the same 2D position
unless they are in opposite layers, intralayer interactions cannot
influence the particles; there is therefore only one type of
interaction generated by the RG flow, interactions between electrons
in opposite layers with the renormalized coupling parameter
$\Gamma_{D}$.  The direct and exchange first order processes in
Fig.~\ref{fig:three} have the values $V_{D}$ and $0$ respectively
where $V_{D}$ is the bare coupling parameter.

The PRG analysis determines how $V_{D}$ is renormalized in a RG
procedure in which fast (high energy) degrees of freedom are
integrated out and the fermion fields of the slow (low energy)
degrees of freedom are rescaled to leave the free-particle action
invariant.  The effective interaction $\Gamma_{D}$ is altered by
coupling between low and high energy degrees of freedom.  At one
loop level this interaction is described\cite{RGShankar} by the
three higher order diagrams labeled ZS, ZS', and BCS in
Fig.~\ref{fig:three}. The internal loops in these diagrams are
summed over the high-energy labels. In the case of 1DES the ZS loop
vanishes and the ZS' and BCS diagrams cancel, implying that the
interaction strengths do not flow to large values and that neither
the CDW repulsive interaction nor the BCS attractive interaction
instabilities predicted by mean-field theory survive the quantum
fluctuations they neglect. The key message of this paper is
summarized by two observations about the properties of these
one-loop diagrams in the bilayer graphene case; i) the
particle-particle (BCS) and particle-hole (ZS, ZS') loops have the
same logarithmic divergences as in the 1DES case in spite of the
larger space dimension and ii) the ZS loop, which vanishes in the
1DES case, is finite in the bilayer graphene case and the BCS loop
vanishes instead.  Both of these changes are due to a layer
pseudospin triplet contribution to the single-particle Green's
function as we explain below.  The net result is that interactions
flow to strong coupling even more strongly than in the mean-field
approximation. The following paragraphs outline key steps in the
calculations which support these conclusions.

An elementary calculation shows that the single-particle Matsubara
Green's function corresponding to the Hamiltonian in
Eq.~(\ref{eq:band}) is
\begin{eqnarray}
{\mathscr G}({\bm k}, i\omega_n)= \left(
\begin{array}{cc}
{\mathscr G}_{\rm s}({\bm k}, i\omega_n) & - {\mathscr G}_{\rm t}({\bm k}, i\omega_n) e^{-iJ\phi_{\bm k}} \\
- {\mathscr G}_{\rm t}({\bm k}, i\omega_n) e^{iJ\phi_{\bm k}} & {\mathscr G}_{\rm s}({\bm k}, i\omega_n) \\
\end{array}
\right)
\end{eqnarray}
where
\begin{equation}
{\mathscr G}_{{\rm s}, {\rm t}}({\bm k}, i\omega_n)\equiv{1\over
2}\left({1\over i\omega_n-\omega_{{\bm k}}}\pm{1\over i\omega_n
+\omega_{\bm k}}\right),
\end{equation}
and $\hbar \omega_{\bm k}=\xi_{\bm k}=\hbar k^2/2m^*$.  The
pseudospin-singlet component of the Green's function ${\mathscr
G}_{\rm s}$, which is diagonal in layer index, changes sign under
frequency inversion whereas the triplet component ${\mathscr G}_{\rm
t}$, which is off-diagonal, is invariant.

The loop diagrams are evaluated by summing the product of two
Green's functions (corresponding to the two arms of the Feynman
diagram loops) over momentum and frequency.  The frequency sums are
standard and yield ($\beta = (k_{\rm B} T)^{-1}$)
\begin{eqnarray}
\label{eq:freq_sum_gf} &&{1\over \beta\hbar^2}\sum_{\omega_n}
{\mathscr G}_{{\rm s}, {\rm t}}^2(i\omega_n) = \mp{\tanh (\beta
\xi_{\bm k}/2) \over 4 \xi_{\bm k} }\mathop {\longrightarrow}
\limits_{T \to 0} \mp{1\over 4 \xi_{\bm k}}\nonumber\\
&&{1\over \beta\hbar^2}\sum_{\omega_n}{\mathscr G}_{\rm
s}(i\omega_n) {\mathscr G}_{\rm t}(i\omega_n) \mathop
{\longrightarrow} \limits_{T \to 0} 0
\end{eqnarray}
where ${\bm k}$ is the momentum label shared by the Green's
functions. Note that the singlet-triplet product sum vanishes in the
low-temperature limit in which we are interested.  Each loop diagram
is multiplied by appropriate interaction constants (discussed below)
and then integrated over high energy momentum labels up to the
massive chiral fermion model's ultraviolet cutoff $\Lambda$:
\begin{equation}
\int_{\Lambda/ s <k<\Lambda} \frac{d^2 {\bm k}}{(2\pi)^2} \;
\frac{\tanh(\beta \xi_{\bm k}/2)}{4\xi_{\bm k}} \; \mathop
{\longrightarrow} \limits_{T \to 0} \;\frac{1}{2}  {\nu_0} \,
\ln(s)\,.
\end{equation}
where $\nu_0=m^*/2\pi\hbar^2$ is the graphene bilayer
density-of-states. Because $\omega_{\bm k} \propto k^2$, this
integral grows logarithmically when the high-energy cut-off is
scaled down by a factor of $s$ in the RG transformation, exactly
like 1DES. This rather surprising property of bilayer graphene is
directly related to its unusual band structure with Fermi points
rather than Fermi lines and quadratic rather than linear dispersion.

The key differences between bilayer graphene and the 1DES appear
upon identifying the coupling factors which are attached to the loop
diagrams.  The external legs in the scattering function Feynman
diagrams (Fig.~\ref{fig:three}) are labeled by layer index ($T =$
top layer and $B = $ bottom layer) in bilayer graphene.  The
corresponding labels for the 1DES are chirality ($R =$ right-going
and $L =$ left going); we call these interaction labels when we
refer to the two cases generically.  Since only opposite layer
interactions are relevant, all scattering functions have two
incoming particles with opposite layer labels and two outgoing
particles with opposite layer labels.  In the ZS loop, at the upper
vertex the incoming and outgoing $T$ particles induce a $B$
particle-hole pair in the loop while the incoming and outgoing $B$
particles at the lower vertex induce a $T$ particle-hole pair. Since
the particle-hole pairs must annihilate each other, there is a
contribution only if the single-particle Green's function is
off-diagonal in interaction labels.  This loop can be thought of as
screening $V_{D}$; the sign of the screening contribution is
opposite to normal, enhancing the bare interlayer interaction,
because the polarization loop involves layer pseudospin triplet
propagation.  (See Eq.~(\ref{eq:freq_sum_gf}).) This contribution is
absent in the 1DES case because propagation is always diagonal in
interaction labels. The ZS' channel corresponds to repeated
interaction between a $T$ particle and a $B$ hole.  This loop
diagram involves only particle-propagation that is diagonal in
interaction labels and its evaluation in the graphene bilayer case
therefore closely follows the 1DES calculation.  This is the channel
responsible for the 1DES mean-field CDW instability in which
coherence is established between $R$ and $L$ particles. In both
graphene bilayer and 1DES cases it has the effect of enhancing
repulsive interactions.  The BCS loop corresponds to repeated
interaction between the two incoming particles.  In the 1DES case
the contribution from this loop which enhances attractive
interactions, cancels the ZS' contribution, leading to marginal
interactions and Luttinger liquid behavior. In the graphene bilayer
case however, there is an additional contribution to the BCS loop
contribution in which the incoming $T$ and $B$ particles both change
interaction labels before the second interaction.  This contribution
is possible because of the triplet component of the particle
propagation and, in light of Eq.~(\ref{eq:freq_sum_gf}), gives a BCS
loop contribution with a sign opposite to the conventional one.
Summing both terms, it follows that the BCS loop contribution is
absent in the graphene bilayer case.  These results are summarized
in Table~\ref{table:one} and imply that at one loop level
\begin{equation}
\Gamma_D \simeq \frac{V_D}{1-V_{D}\nu_0 \ln(s)}.
\end{equation}
The interaction strength diverges when $V_{D}\,\nu_0 =1/ \ln(s)$, at
half\footnotemark[2]\footnotetext[2]{Mean-field theory is equivalent
to a single-loop PRG calculation in which only one particle-hole
channel is retained. For the drawing conventions of
Fig.~\ref{fig:three} the susceptibility which diverges at the
pseudospin ferromagnet phase boundary is obtained by closing the
scattering function with $\tau^{z}$ vertices at top and bottom, so
the appropriate particle-hole channel is the ZS channel
not the ZS' channel.} the mean-field theory critical interaction strength.
Taking guidance from the mean-field theory\cite{min_prb_2008}, the
strong coupling state is likely a pseudospin ferromagnet which has
an energy gap and spontaneous charge transfer between layers.
\begin{table}
\caption{Summary of contrasting the contributions (in units of the
related density-of-states) of the three one-loop diagrams in 1DES
and graphene bilayer cases} \centering
\begin{tabular}{| c | c | c | c | c |}

\hline

diagrams & ZS & ZS' & BCS & one-loop\\

\hline

1DES & $0$ & $u^2\,\ln(s)$ & $-u^2\,\ln(s)$ & 0\\

\hline

graphene bilayer & $\frac{1}{2}\,\Gamma_D^2\,\ln(s)$ & $\frac{1}{2}\,\Gamma_D^2\,\ln(s)$ & $0$ & $\Gamma_D^2\,\ln(s)$\\

\hline

\end{tabular}
\label{table:one}
\end{table}

A number of real-world complications which have to be recognized in
assessing the experimental implications of these results.  First of
all, electrons in real graphene bilayers carry spin and valley as
well as layer pseudospin degrees of freedom.  This substantially
complicates the PRG analysis since many different types of
interactions are generated by the RG flow.  When one spin degree of
freedom is considered three types of interactions have to be
recognized, $\Gamma_S$, $\Gamma_D$, and $\Gamma_X$. $\Gamma_S$
couples electrons with the same flavor and $\Gamma_D$ electrons with
different flavors, while $\Gamma_X$ is the exchange counterpart of
$\Gamma_D$. The bare value of the exchange part of $\Gamma_D$ (the
direct part of $\Gamma_X$) is of course zero since the Coulomb
interaction is flavor independent, but higher order contributions
are nonzero if triplet electron propagation is allowed.
Correspondingly, when both spin and valley degrees of freedom are
acknowledged the interaction parameters are $\Gamma_{SSD}$,
$\Gamma_{SDS}$, $\Gamma_{SDD}$, $\Gamma_{DSS}$, $\Gamma_{DSD}$,
$\Gamma_{DDS}$, $\Gamma_{DDD}$, $\Gamma_{XSD}$, $\Gamma_{XDS}$, and
$\Gamma_{XDD}$ (see supplementary material) . The labels refer from
left to right to sublattice pseudospin, real spin, and valley
degrees of freedom. $\Gamma_{SSS}$ is absent due to Pauli exclusion
principle while $\Gamma_{XSS}$ is just $-\Gamma_{DSS}$ due to the
fermionic antisymmetry between outgoing particles. The one loop flow
of these ten interaction parameters is described in the
supplementary material.  We find that the renormalized interactions
diverge near $V_{D}\,\nu_0 \simeq 0.6/\ln(s)$. The instability
tendency is therefore somewhat enhanced by the spin and the valley
degrees of freedom.

Next we must recognize that the massive chiral fermion model applies
only between trigonal-warping and interlayer hopping energy scales.
Appropriate values for $\ln(s)$ in the RG flows are therefore at
most around $2.3$ (see supplementary material) in bilayer graphene,
compared to the unlimited values of the chiral-fermion model. When
combined with estimates of the bare interaction strengths (see
supplementary material), this limit on interaction flow suggests
that spontaneous gaps are likely in suspended bilayer graphene
samples, and much less likely for graphene bilayers on substrates.

Gaps do appear in bilayer graphene even when electron-electron
interactions are neglected, provided that an external potential
difference $V$ is applied between the layers.  The potential adds a
single-particle term $-V\tau^{z}/2$ to the single-particle
Hamiltonian, breaks inversion symmetry, and transfers
charge\cite{ohta_science_2006,castro_prl_2007,SPBilayer_1,SPBilayer_2}
between layers. This interesting property is in fact the basis of
one strategy currently being explored in the effort to make useful
electronic devices\cite{oostinga_naturemat_2008,jens_MM_2009} out of
graphene 2DES's. Even if gaps do not appear spontaneously in real
bilayer graphene samples, it is clear from the present work that
interesting many-body physics beyond that captured by commonly used
electronic-structure-theory approximations (LDA or GGA
approximations for example), must play at least a quantitative role
in determining gap grown with $V$.  As graphene bilayer sample
quality improves, we expect that it will be possible to explore this
physics experimentally with ARPES, tunneling, and transport probes.

\acknowledgements
 This work has been supported by the Welch Foundation and by the National Science Foundation under grant
 DMR-0606489. AHM gratefully acknowledges helpful discussion with R.
 Shankar and I. Affleck.


\end{document}


\title{Supplementary Information for ``Electron-electron interactions in graphene bilayers''}


\author{Fan Zhang$^{1}$, Hongki Min$^{1}$, Marco Polini$^{2}$ and A.H. MacDonald$^{1}$}




\affiliation{
$^{1}$Department of Physics, University of Texas at
Austin, Austin TX 78712, USA \\
$^{2}$NEST-CNR-INFM and Scuola Normale Superiore, I-56126 Pisa,
Italy}

\maketitle

In this supplementary information, we summarize the Green's function
technique, detail the diagrammatic perturbation theory calculations,
and explain the many interaction channels generated by the RG flows
when spin and valley degrees of freedom are included.  We also
discuss the bare interactions and the important trigonal-warping
effect.

\section{Green's functions and Frequency Sums}
The low-energy effective Hamiltonian of a graphene bilayer is a momentum ${\bm q}$-dependent
$2\times 2$ matrix of the
following form:
\begin{equation}
{\mathscr H}= - {\bm B}\cdot {\bm \tau}~,
\end{equation}
where ${\bm \tau}$ is the Pauli matrix vector, $|{\bm B}|=\xi_{\bm
q}=\hbar \omega_{\bm q} = \hbar^2q^2/2m^*$, and the orientation
angle of ${\bm B}$ is twice the orientation angle of ${\bm q}$ ($\phi_{\bm q}$).
(This property is often expressed by saying that the model has layer
pseudospin chirality~\cite{McCann,min_2008_b} $J$=2.) The pseudospin
degree of freedom upon which the operator ${\bm \tau}$ acts
represents the layer in which an electron resides.  Eigenstates of
${\mathscr H}$ are coherent linear combinations of amplitudes in
both layers. The finite temperature Green's function at Matsubara
fermion frequency $i\omega_n$ and wavevector ${\bm q}$ is given by
\begin{equation}
{\mathscr G}(i\omega_n)= [i\omega_n- {\mathscr
H}/\hbar]^{-1} = {\mathscr G}_{\rm s}(i\omega_n)+{\mathscr
G}_{\rm t}(i\omega_n) \; {\bm n} \cdot {\bm \tau}\,
\end{equation}
where
\begin{equation}
{\mathscr
G}_{{\rm s}, {\rm t}}(i\omega_n)\equiv{1\over 2}\left({1\over
i\omega_n-\omega_{\bm q}}\pm{1\over i\omega_n+\omega_{\bm
q}}\right),
\end{equation}
${\bm n}\equiv - {\bm  B}/|{\bm B}|$, the ``${\rm s}$" and ``${\rm t}$"
subscripts denote singlet and triplet contributions respectively,
and we have chosen $\mu=0$ to address the properties of an
electrically neutral bilayer. Note that
${\mathscr G}_{\rm s}(-i\omega_n) = -{\mathscr G}_{\rm s}(i\omega_n)$
whereas
${\mathscr G}_{\rm t}(-i\omega_n) = {\mathscr G}_{\rm t}(i\omega_n)$.  Using the property
that $\mathbf{n} = -\,(\cos(2\phi_{\bm q}),\sin(2\phi_{\bm q}),0)$
we obtain the following explicit form for the $2\times2$
Green's function matrix:
\begin{equation}
\label{eq:phasefactor} {\mathscr G}(i\omega_n)= \left(
\begin{array}{cc}
{\mathscr G}_{\rm s}(i\omega_n) & -{\mathscr G}_{\rm t}(i\omega_n) e^{-2 i \phi_{\bm q}} \\
-{\mathscr G}_{\rm t}(i\omega_n) e^{2 i \phi_{\bm q}} & {\mathscr G}_{\rm s}(i\omega_n)  \\
\end{array}
\right).
\end{equation}
The off diagonal triplet component captures processes in which
free electrons propagate from one layer to the other.
Evaluation of the loop diagrams which appear in the RG calculation described in the main text requires
frequency sums to be performed for products of two Green's functions.  These are given by
\begin{eqnarray}
\label{eq:freq_sum_gf} {1\over \beta\hbar^2}\sum_{\omega_n}
{\mathscr G}_{{\rm s}, {\rm t}}^2(i\omega_n) =
\mp{\tanh (\beta \xi_{\bm q}/2) \over 4 \xi_{\bm q} }\mathop
{\longrightarrow} \limits_{T \to 0} \mp{1\over 4 \xi_{\bm q}}\,,
\quad {1\over \beta\hbar^2}\sum_{\omega_n} {\mathscr G}_{\rm s}(i\omega_n)
{\mathscr G}_{\rm t}(i\omega_n) \mathop {\longrightarrow} \limits_{T \to 0} 0\,.
\end{eqnarray}

\section{Spins Pseudospins and Distinct Interaction Parameters}
\begin{figure}[htbp]
\centering \scalebox{0.7} {\includegraphics*
[1.9in,7.3in][8.2in,9.55in] {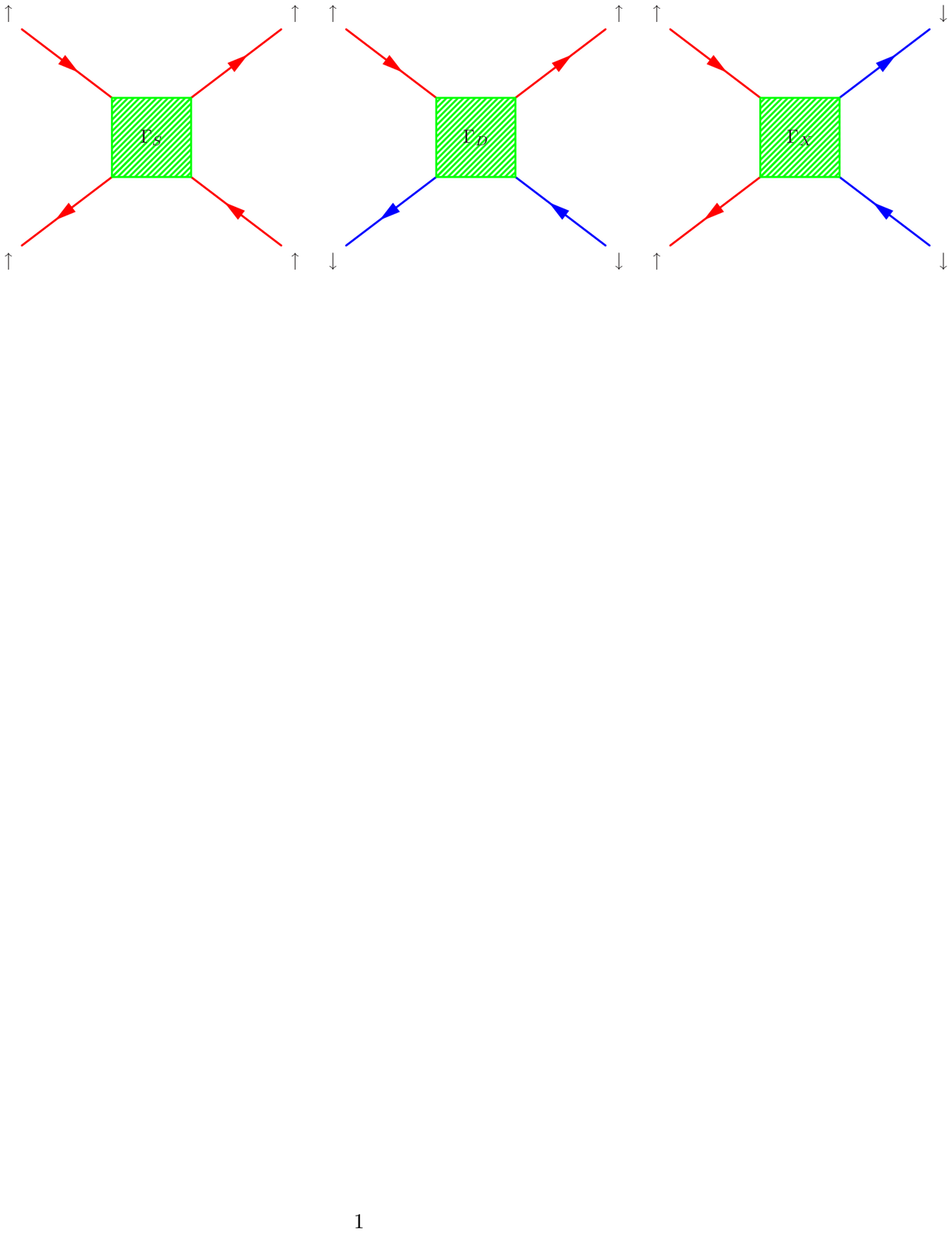}} \caption{\label{fig:gamma}\bf
Electron-electron scattering processes for a system with one
pseudospin-$1/2$ degree of freedom.}
\end{figure}
In the low-energy continuum model of bilayer graphene electrons
carry spin, and both layer and valley pseudospin labels. In a
scattering event, both the two incoming and two outgoing particles
can therefore have one of eight labels and the general scattering
function therefore has $8^4$ possible low-energy long-wavelength
values.  The number of distinct coupling constants in the RG flow
equations is much smaller, however, because many values are zero and
others are related to each other by symmetry. One simplification is
that interactions conserve spin, and both layer and valley
pseudospin, at each vertex.  Interactions are however dependent on
whether the interacting particles are in the same ($S$) or in
different ($D$) layers.  The internal loops in the perturbative RG
calculation contain two fermion propagator (Green's function) lines.
These propagators conserve both spin and valley pseudospin, but as
we have seen above, not the layer pseudospin.  It is clear then that
the incoming and outgoing total spin must be preserved for real spin
and for the valley pseudospin, but the layer pseudospin case
requires a more elaborate consideration.
From Eq.~(\ref{eq:phasefactor}) we see that a phase factor $e^{\pm
2i \phi_{\bm q} }$ is gained when the propagator transfers electrons
between layer index with the $+$ for top to bottom evolution and the
$-$ for bottom to top. Unless these transfers enter an equal number
of times in each direction, the integrand in a Feynman diagram will
contain a net phase factor related to chirality and vanish under
momentum integration. The total layer pseudospin is therefore also
conserved in collisions.

In identifying distinct coupling constants, we start with the
simplest case in which the valley and spin labels are absent. There
are then three possibilities, as illustrated in
Fig.~\ref{fig:gamma}. When the two incoming pseudospins are
parallel ($\Gamma_S$ in Fig.~\ref{fig:gamma}), the outgoing
pseudospins must also be parallel.  Because of Fermi statistics
interchanging the outgoing lines in $\Gamma_S$ changes the
diagrams's sign.  Since the diagram is invariant under this
operation, it must vanish. The second possibility is opposite
incoming pseudospins, which requires opposite outgoing pseudospins
in one of the two configurations labelled by $\Gamma_D$ and
$\Gamma_X$ in Fig.~\ref{fig:gamma}.   In this case Fermi
statistics implies that $\Gamma_D=-\Gamma_X$. It follows that the
only distinct interaction parameter is $\Gamma_D$.

\begin{figure}[htbp]
\centering \scalebox{1} {\includegraphics*
[1.8in,8.1in][7.8in,9.40in] {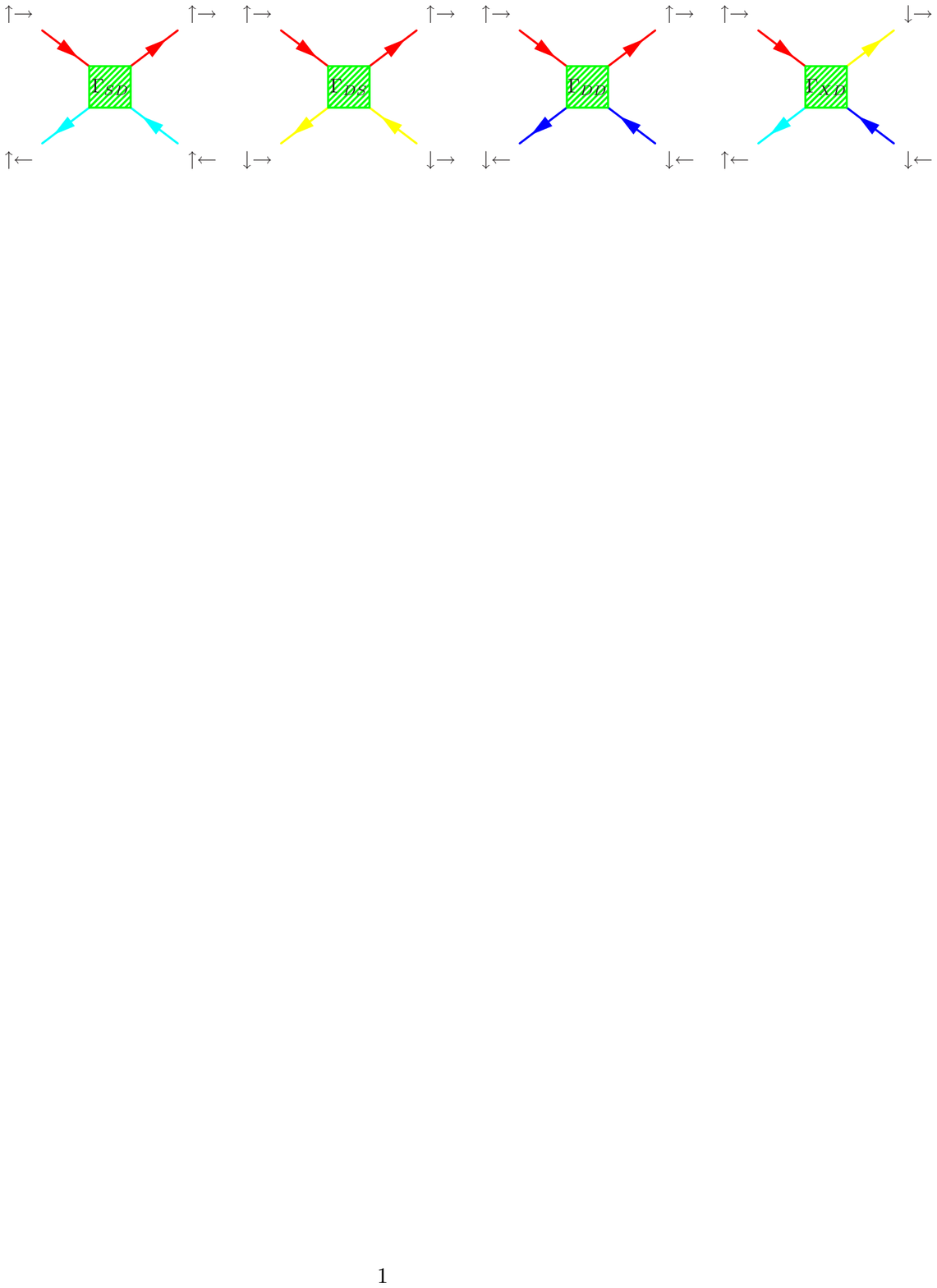}} \caption{\label{fig:DXXD}
{\bf Distinct interaction channels for systems with two
pseudospin-${1/2}$ degrees of freedom.} In this figure the first
spin is denoted by $|\uparrow \rangle$ or $|\downarrow \rangle$
while the second by $|\rightarrow \rangle$ or $|\leftarrow \rangle$.
The one-dimensional electron gas system can be viewed as being in this class if the
chirality index is regarded as a pseudospin.}
\end{figure}

If more than one pseudospin is present, we have to recognize more
separate interacting processes.  For example, for systems with two
relevant pseudospins, the interaction parameters can be labeled in
the same way as in Fig.~\ref{fig:gamma} but by doublets which
account for the different pseudospins separately.  Two pseudospin
interactions might include $\Gamma_{SD}$, $\Gamma_{DS}$,
$\Gamma_{DD}$ and $\Gamma_{XD}$ (see Fig.~\ref{fig:DXXD}) for
example. In models for which propagators and interactions preserve
all pseudospin labels, we would have $\Gamma_{XD}=0$ since all
pseudospin flavors are preserved along each fermion line.  For
graphene bilayers, however, we must keep $\Gamma_{XD}\neq0$ because
the layer pseudospin has triplet propagation.  Following this line
of argument, we conclude that in graphene bilayers, with its three
different pseudospins, there are ten distinct non-zero interaction
parameters: $\Gamma_{SSD}$, $\Gamma_{SDS}$, $\Gamma_{SDD}$,
$\Gamma_{DSS}$, $\Gamma_{DSD}$, $\Gamma_{DDS}$, $\Gamma_{DDD}$,
$\Gamma_{XSD}$, $\Gamma_{XDS}$ and $\Gamma_{XDD}$, where the first
label refers to layer pseudospin, and the following labels to real
spin and valley.

\section{PRG analysis for spinless and valleyless graphene bilayers}
The one-loop correction to the interlayer interaction $\Gamma_D$
can be calculated using the following ZS, ZS' and BCS
diagrams~\cite{shankar1994}.
Since $\Gamma_D$ (with bare value $V_D$) is the only interaction
parameter as we explained previously, the propagator labels of the
internal loops must be distributed accordingly. We label the
external legs in the scattering diagrams by their layer indices
($T =$ top layer and $B =$ bottom layer). The corresponding labels for the
1DES are chirality ($R =$ right-going and $L =$ left going).
\begin{figure}[htbp]
\centering \scalebox{0.8}
{\includegraphics*[2.1in,7.3in][4.4in,9.55in]{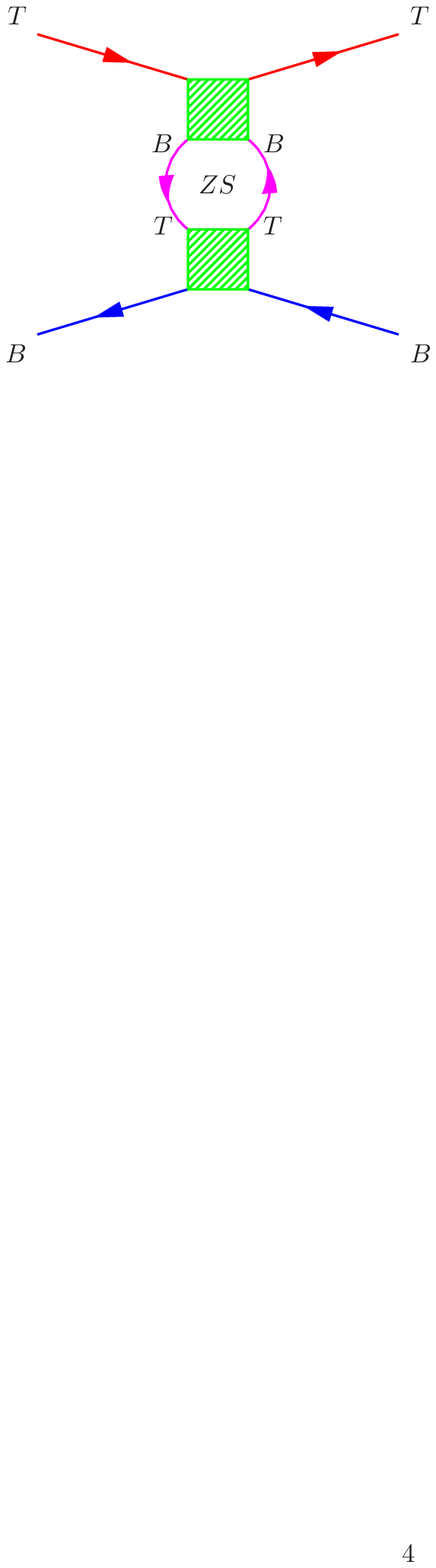}}
\caption{\label{fig:zs}\bf ZS loop correction in the one-loop
perturbative RG calculation.}
\end{figure}

In the ZS loop shown in Fig.~\ref{fig:zs}, at the upper vertex the
incoming and the outgoing $T$ particles induce a $B$ particle-hole
pair in the loop while the incoming and outgoing $B$ particles at
the lower vertex induce a $T$ particle-hole pair.  The corresponding
labels in the 1DES case are $L$, $R$ for left and right chirality.
This correction survives in the bilayer case only because the
single-particle Green's function has a triplet contribution [see
Eq.~(\ref{eq:freq_sum_gf})] which is off-diagonal in layer index.
The ZS contribution is absent in the 1DES
case~\cite{shankar1994,giamarchibook} because propagation is always
diagonal in interaction labels. Here we find
\begin{eqnarray}
\Gamma_{D}^{\rm ZS}&=&{\Gamma_D^2\over \beta\hbar^2}\int {d^2 {\bm q}\over
(2\pi)^2}\,\sum_{\omega_n} {\mathscr{G}}_{\rm t}^2(\bm{q},i\omega_n)
=\Gamma_D^2\int {d^2 {\bm q}\over (2\pi)^2}\, {\tanh (\beta \xi_{\bm q}/2)
\over 4 \xi_{\bm q} }={1\over 2}\, \Gamma_D^2\,\nu_0\ln (s)\,,
\end{eqnarray}
where $\nu_0=m^*/2\pi\hbar^2$ is the graphene bilayer
density-of-states (per spin and valley) and the integral is carried out in the momentum
shell $\Lambda/ s <q<\Lambda$.  In the zero temperature limit
\begin{equation}
\int_{\Lambda/ s <q<\Lambda} \frac{d^2 {\bm q}}{(2\pi)^2} \;
\frac{\tanh(\beta \xi_{\bm q}/2)}{4\xi_{\bm q}} \; \mathop {\longrightarrow}
\limits_{T \to 0} \;\frac{1}{2}  {\nu_0} \, \ln(s)\,.
\end{equation}

\begin{figure}[htbp]
\centering \scalebox{0.8}
{\includegraphics*[2.1in,7.3in][4.4in,9.55in]{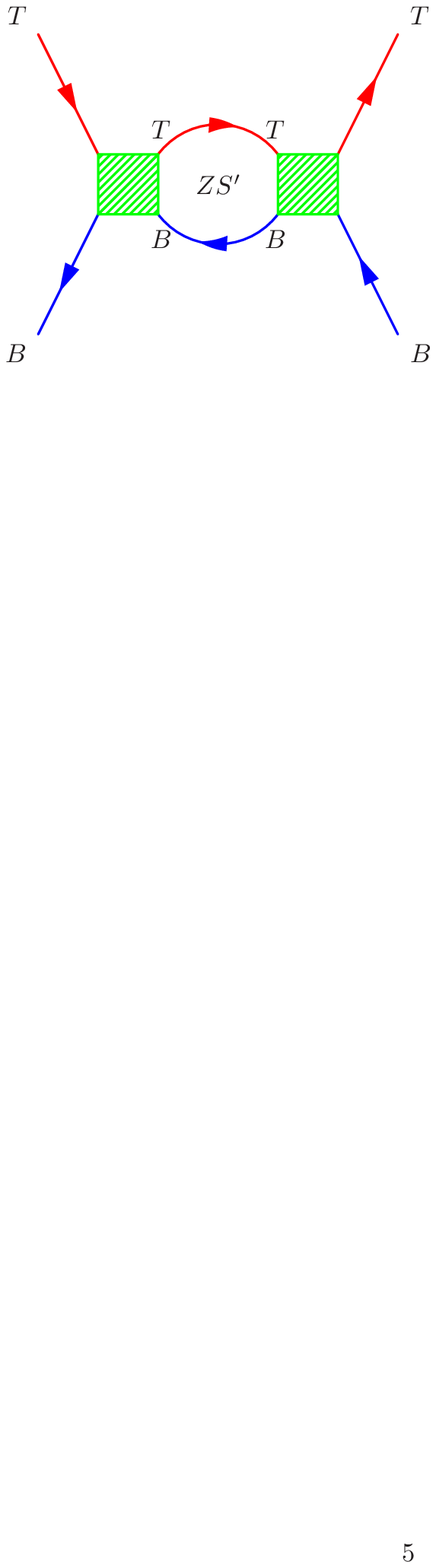}}
\caption{\label{fig:zs'}\bf ZS' loop correction in the one-loop
perturbative RG.}
\end{figure}

The ZS' channel shown in Fig.~\ref{fig:zs'} corresponds to
repeated interaction between a $T$ particle and a $B$ hole.  This loop
diagram involves only particle-propagation that is diagonal in
interaction labels; its evaluation for the 1DES and graphene
bilayers correspond closely. This is the channel responsible for the
1DES mean-field CDW instability~\cite{shankar1994} in  which
coherence is established between $R$ and $L$
particles~\cite{giamarchibook}.  In both cases it has the effect of
enhancing repulsive interactions. We find
\begin{eqnarray}
\Gamma_{D}^{{\rm ZS'}}&=&-{\Gamma_D^2\over \beta\hbar^2}\int {d^2 {\bm q}\over
(2\pi)^2}\,\sum_{\omega_n} {\mathscr{G}}_{\rm s}^2(\bm{q},i\omega_n)
=\Gamma_D^2\int {d^2 {\bm q}\over (2\pi)^2}\, {\tanh (\beta \xi_{\bm q}/2)
\over 4 \xi_{\bm q} }={1\over 2}\, \Gamma_D^2\,\nu_0\ln (s)\,.
\end{eqnarray}

\begin{figure}[htbp]
\centering \scalebox{0.8}
{\includegraphics*[2.1in,7.3in][6.9in,9.55in]{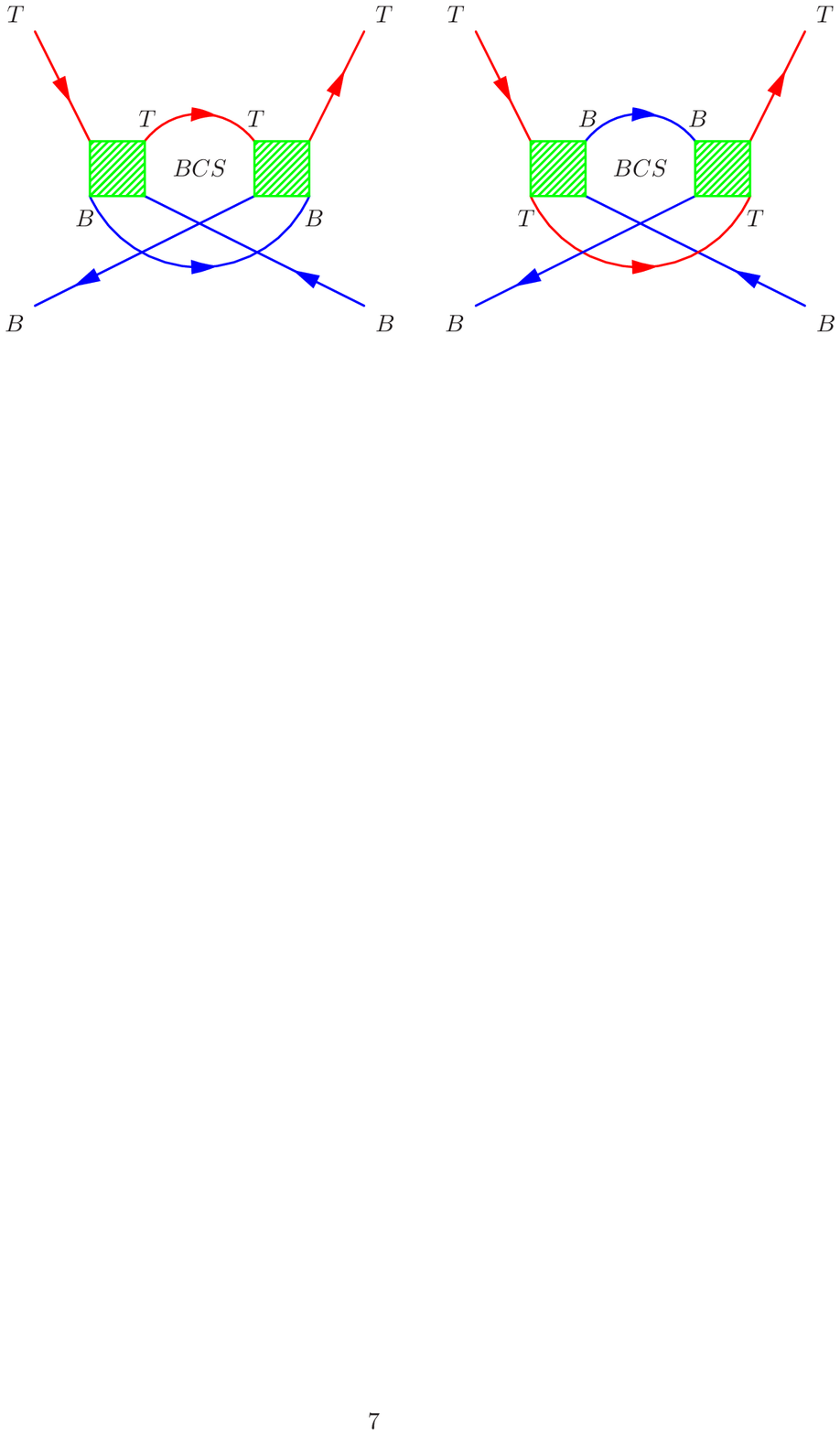}}
\caption{\label{fig:bcs1}\bf BCS(particle-particle) loop correction
for singlet propagation in the one-loop perturbative RG
calculation.}
\end{figure}

The BCS loop corresponds to repeated interaction between the two
incoming particles.  In the 1DES case the contribution from this
loop (see Fig.~\ref{fig:bcs1}) cancels the ZS'
contribution~\cite{shankar1994}, leading to marginal interactions and
Luttinger liquid behavior. This same kind of BCS correction for
graphene bilayer reads
\begin{eqnarray}
\label{eq:bcs1} \Gamma_{D}^{{\rm BCS}_1}=&-&{1\over 2}{\Gamma_D^2\over
\beta\hbar^2}\int {d^2 {\bm q}\over (2\pi)^2}\,\sum_{\omega_n}
{\mathscr{G}}_{\rm s}(\bm{q},i\omega_n){\mathscr{G}}_{\rm s}(-\bm{q},-i\omega_n)\nonumber\\
&-&{1\over 2}{(-\Gamma_D)^2\over \beta\hbar^2}\int {d^2 {\bm q}\over
(2\pi)^2}\,\sum_{\omega_n} {\mathscr{G}}_{\rm s}(\bm{q},i\omega_n){\mathscr{G}}_{\rm s}(-\bm{q},-i\omega_n)\nonumber\\
=&-&{1\over 2}\Gamma_D^2\,\nu_0\ln (s)\,.
\end{eqnarray}
In the graphene bilayer case, however, there is an additional
contribution (see Fig.~\ref{fig:bcs2}) to the BCS loop contribution
in which the incoming $T$ and $B$ particles both change layer labels
before the second interaction.  This contribution is possible
because of the triplet layer pseudospin propagation and, in light of
Eq.~(\ref{eq:freq_sum_gf}), gives a BCS contribution of opposite
sign to the normal contribution:
\begin{figure}[htbp]
\centering \scalebox{0.8} {\includegraphics*
[2.1in,7.3in][6.9in,9.55in] {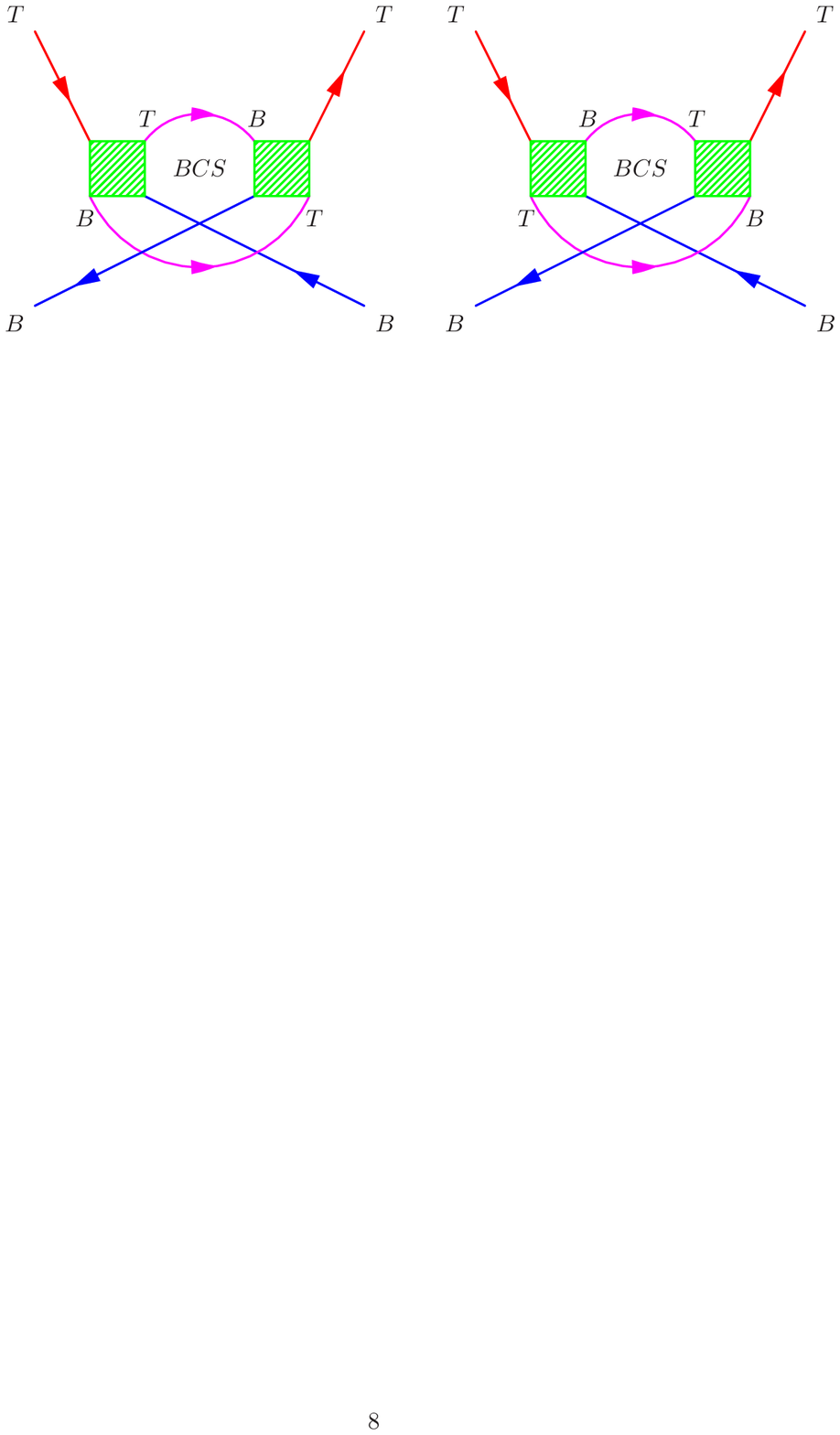}}
\caption{\label{fig:bcs2}\bf BCS(particle-particle) loop correction
for triplet propagation in the one-loop perturbative RG
calculation.}
\end{figure}
\begin{eqnarray}
\Gamma_{D}^{{\rm BCS}_2}&=&-{1\over 2}{\Gamma_D(-\Gamma_D)\over
\beta\hbar^2}\int {d^2 {\bm q}\over (2\pi)^2}\,\sum_{\omega_n}
{\mathscr{G}}_{\rm t}(\bm{q},i\omega_n){\mathscr{G}}_{\rm t}(-\bm{q},-i\omega_n)\nonumber\\
&&-{1\over 2}{(-\Gamma_D)\Gamma_D\over \beta\hbar^2}\int {d^2 {\bm q}\over
(2\pi)^2}\,\sum_{\omega_n}
{\mathscr{G}}_{\rm t}(\bm{q},i\omega_n){\mathscr{G}}_{\rm t}(-\bm{q},-i\omega_n)\nonumber\\
&=&{1\over 2}\Gamma_D^2\,\nu_0\ln (s) \,.
\end{eqnarray}
It follows that the BCS loop contribution is absent in the graphene
bilayer case because
\begin{eqnarray}
\Gamma_D^{\rm BCS}=\Gamma_D^{{\rm BCS}_1}+\Gamma_D^{{\rm BCS}_2}=0\,.
\end{eqnarray}

Therefore, at one-loop level, the renormalization of interlayer
interaction is
\begin{eqnarray}
\Gamma_D^{{\rm one-loop}}=\Gamma_D^{\rm ZS}+\Gamma_D^{{\rm ZS'}}+\Gamma_D^{\rm BCS}=\Gamma_D^2\,\nu_0\ln(s)\,,
\end{eqnarray}
which leads to the RG flow equation
\begin{eqnarray}
\frac{{\emph d}\,\Gamma_{D}}{\nu_0\,{\emph d}\ln(s)}=\Gamma_D^2\,.
\end{eqnarray}
Combined with the bare interlayer interaction $V_D$ and integrating
the flow equation we obtain that
\begin{equation}
\Gamma_D = \frac{V_D}{1-V_{D}\nu_0 \ln(s)}
\end{equation}
which diverges at the point $V_{D}\,\nu_0 =1/ \ln(s)$.

For the Feynman diagrams drawing conventions we have chosen, the
interaction correction to the layer pseudospin response function
$\chi_{zz}$, which diverges at the pseudospin ferromagnet phase
boundary, is obtained by closing the scattering function with a
$\tau_{z}$ vertex at top and bottom.  The $\tau_{z}$ operator
measures the charge difference between $T$ and $B$ layers.  Because
it is an effective single-particle theory, fermion mean-field
theory~\cite{min_prb_2008} corresponds to response function diagrams
with at most a single particle-hole pair.  It follows that
mean-field theory is equivalent to a single-loop PRG calculation in
which the BCS and ZS' channels are neglected and only the ZS
channels is retained.  In mean-field theory~\cite{min_prb_2008} the
ideal graphene bilayer has an instability to a state in which charge
is spontaneously transferred between the layers which is signalled
by the divergence of $\chi_{zz}$.  The PRG analysis demonstrates
that the mean-field theory instability is enhanced by reinforcing
ZS' channel contribution.

\section{PRG analysis for spinful and valleyful graphene bilayers}
As discussed at the end of section II, there are ten interaction
parameters for the spinful and valleyful case.  The one-loop flow
equations are derived in the same way as in the spinless valleyless
case, except for the necessity of keeping track of the many-possible
configurations of the end labels on the loop propagators.  We find
that
\begin{eqnarray}
\label{eq:full_rg_flow}
\frac{{\emph d}\,\Gamma_{SSD}}{\nu_0\,{\emph d}\ln(s)}&=&-\frac{1}{2}\Gamma_{SSD}^2-\Gamma_{DSS}(\Gamma_{DSD}-\Gamma_{SSD})-(\Gamma_{DDS}-\Gamma_{SDS})(\Gamma_{DDD}-\Gamma_{SDD})+\frac{1}{2}(\Gamma_{XSD}-\Gamma_{SSD})^2\nonumber\\
\frac{{\emph d}\,\Gamma_{SDS}}{\nu_0\,{\emph d}\ln(s)}&=&-\frac{1}{2}\Gamma_{SDS}^2-\Gamma_{DSS}(\Gamma_{DDS}-\Gamma_{SDS})-(\Gamma_{DSD}-\Gamma_{SSD})(\Gamma_{DDD}-\Gamma_{SDD})+\frac{1}{2}(\Gamma_{XDS}-\Gamma_{SDS})^2\nonumber\\
\frac{{\emph d}\,\Gamma_{SDD}}{\nu_0\,{\emph d}\ln(s)}&=&-\frac{1}{2}\Gamma_{SDD}^2-\Gamma_{DSS}(\Gamma_{DDD}-\Gamma_{SDD})-(\Gamma_{DSD}-\Gamma_{SSD})(\Gamma_{DDS}-\Gamma_{SDS})+\frac{1}{2}(\Gamma_{XDD}-\Gamma_{SDD})^2\nonumber\\
\nonumber\\
\frac{{\emph d}\,\Gamma_{DSS}}{\nu_0\,{\emph d}\ln(s)}&=&\Gamma_{DSS}^2+\frac{1}{2}(\Gamma_{DSD}-\Gamma_{SSD})^2+\frac{1}{2}(\Gamma_{DDS}-\Gamma_{SDS})^2+\frac{1}{2}(\Gamma_{DDD}-\Gamma_{SDD})^2
+\frac{1}{2}\Gamma_{XSD}^2+\frac{1}{2}\Gamma_{XDS}^2+\frac{1}{2}\Gamma_{XDD}^2\nonumber\\
\frac{{\emph d}\,\Gamma_{DSD}}{\nu_0\,{\emph d}\ln(s)}&=&\frac{1}{2}\Gamma_{DSD}^2+\Gamma_{DSS}(\Gamma_{DSD}-\Gamma_{SSD})+(\Gamma_{DDD}-\Gamma_{SDD})(\Gamma_{DDS}-\Gamma_{SDS})-\frac{1}{2}(\Gamma_{DSD}+\Gamma_{XSD})^2\nonumber\\
\frac{{\emph d}\,\Gamma_{DDS}}{\nu_0\,{\emph d}\ln(s)}&=&\frac{1}{2}\Gamma_{DDS}^2+\Gamma_{DSS}(\Gamma_{DDS}-\Gamma_{SDS})+(\Gamma_{DDD}-\Gamma_{SDD})(\Gamma_{DSD}-\Gamma_{SSD})-\frac{1}{2}(\Gamma_{DDS}+\Gamma_{XDS})^2\nonumber\\
\frac{{\emph d}\,\Gamma_{DDD}}{\nu_0\,{\emph d}\ln(s)}&=&\frac{1}{2}\Gamma_{DDD}^2+\Gamma_{DSS}(\Gamma_{DDD}-\Gamma_{SDD})+(\Gamma_{DDS}-\Gamma_{SDS})(\Gamma_{DSD}-\Gamma_{SSD})-\frac{1}{2}(\Gamma_{DDD}+\Gamma_{XDD})^2\nonumber\\
\nonumber\\
\frac{{\emph d}\,\Gamma_{XSD}}{\nu_0\,{\emph d}\ln(s)}&=&\Gamma_{DSS}\Gamma_{XSD}-\Gamma_{XDS}\Gamma_{XDD}-\frac{1}{2}(\Gamma_{XSD}-\Gamma_{SSD})^2-\frac{1}{2}(\Gamma_{XSD}+\Gamma_{DSD})^2\nonumber\\
\frac{{\emph d}\,\Gamma_{XDS}}{\nu_0\,{\emph d}\ln(s)}&=&\Gamma_{DSS}\Gamma_{XDS}-\Gamma_{XSD}\Gamma_{XDD}-\frac{1}{2}(\Gamma_{XDS}-\Gamma_{SDS})^2-\frac{1}{2}(\Gamma_{XDS}+\Gamma_{DDS})^2\nonumber\\
\frac{{\emph d}\,\Gamma_{XDD}}{\nu_0\,{\emph
d}\ln(s)}&=&\Gamma_{DSS}\Gamma_{XDD}-\Gamma_{XSD}\Gamma_{XDS}-\frac{1}{2}(\Gamma_{XDD}-\Gamma_{SDD})^2-\frac{1}{2}(\Gamma_{XDD}+\Gamma_{DDD})^2
\end{eqnarray}
\begin{figure}[htbp]
\centering \scalebox{0.40} {\includegraphics* {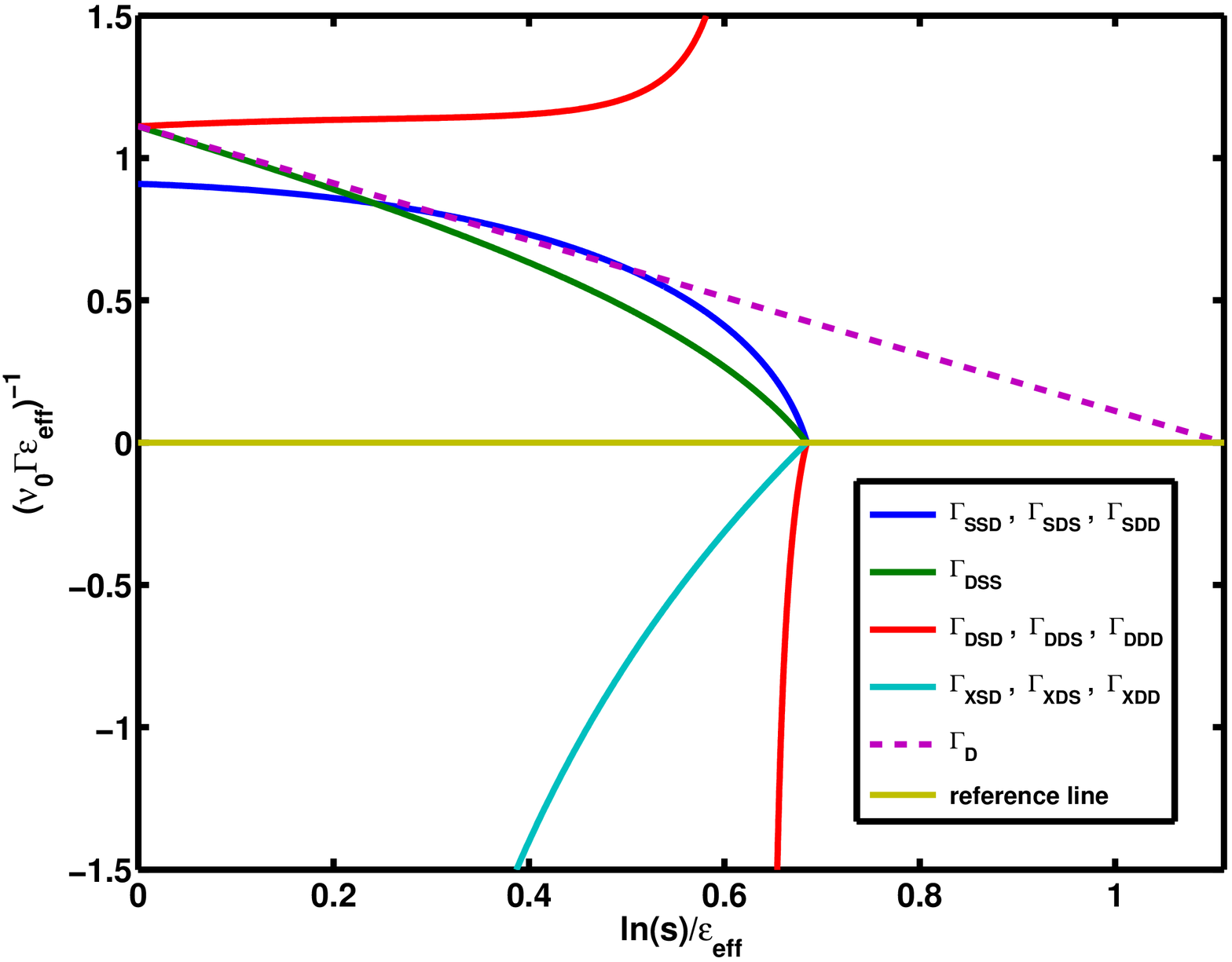}}
\caption{\label{fig:rgflow}{{\bf Renormalized interaction flow.}
This illustration plots the inverse interaction strength
$(\nu_0\Gamma\varepsilon_{\rm eff})^{-1}$ versus the scaling
parameter $\ln(s)/\varepsilon_{\rm eff}$. $\varepsilon_{\rm eff}$ is
the effective dieletric constant of the graphene bilayer and
$\Gamma=\Gamma_{\rm vacuum}/\varepsilon_{\rm eff}$. Interlayer
interaction parameters ${\Gamma_{DSS}}$ (green) and
${\Gamma_{XSD},\Gamma_{XDS},\Gamma_{XDD}}$ (cyan) flow to large
values most quickly. According to this estimate the normal state
becomes unstable for $V_{D}$ larger than $\simeq 0.6/\ln(s)$.}}
\end{figure}

The only fixed point that we have identified is the non-interacting
one. These ten coupled flow equations can be integrated numerically
starting from bare interactions.  In order to represent the property
that same layer interactions will be slightly stronger than
different layer interactions we set the bare interactions values to
 $1.1$,\,$0.9$ and $0$ for
$\nu_0\,V_{SSD}$\,($V_{SDS}$,$V_{SDD}$)\,,$\nu_0\,V_{DSS}$\,($V_{DSD},V_{DDS},V_{DDD}$)
and $\nu_0\,V_{XSD}$\,($V_{XDS},V_{XDD}$), respectively.  (The
motivation for this choice is explained in the next section.) We
find that the interaction parameters flow away from the
non-interacting fixed point and diverge at a finite value of $s$ as
illustrated in Fig.~\ref{fig:rgflow}. The instability criterion
implied by this one-loop PRG calculation is $V_{D}\,\nu_0 \simeq 0.6
/ \ln(s)$.  The instability tendency is therefore enhanced by the
spin and valley degrees of freedom since the criterion was
$V_{D}\,\nu_0 \simeq{{1}/\ln(s)}$ for the spinless and valleyless
case.

\section{Bare interactions and the trigonal-warping effect}
The conclusions which can be drawn from the PRG calculation presented here
are sensitive to the upper and lower momentum and energy
cutoffs, which limit the applicability of the massive chiral fermion model
for bilayer graphene, and to the strength of bare electron-electron scattering amplitudes.
Below we estimate the range of $s$ over which the RG flows discussed above apply,
and the strength of the bare interaction $V_{D}$.
We caution that, given the nature of the PRG calculations, the estimates
presented below should be regarded as qualitative.

In practice the upper cutoff is the interlayer hopping energy
$\gamma_1 \sim 0.4~{\rm eV}$; at higher energies it is essential to
account for two sublattice sites in each layer.  We have in addition
ignored the trigonal-warping part in the full Hamiltonian, due to a
direct hopping process between the low-energy sites which has energy
scale~\cite{rmp,Ando} $\gamma_3 \sim 0.3~{\rm eV}$. Inserting the
expression~\cite{McCann} for the effective mass of the massive chiral
Fermion model we find that the model we have studied is appropriate
for
\begin{equation}
\hbar\,v_{\rm F}\,q\,\frac{\gamma_3}{\gamma_0}\,\leq \, \frac{\hbar^2\,q^2}{2\,m^*} = \frac{\hbar^2\,q^2 v_{\rm F}^2}{\gamma_1} \leq \gamma_1
\end{equation}
where $v_{\rm F} \sim 10^{8}~{\rm cm}/{\rm s}$ is the Fermi velocity
near the Dirac point in the single-layer-graphene continuum model,
and $\gamma_0 \sim 3~{\rm eV}$ is the intralayer near neighbor
hopping energy. It follows that the high energy momentum cutoff
$q_{\rm H} = \gamma_1/\hbar v_{\rm F}$ and that the low energy
momentum cutoff $q_{\rm L} = (\gamma_3 \gamma_1/\gamma_0)/\hbar
v_{\rm F}$,
which gives the maximum value of the scaling parameter $\ln(s)$.
Using accepted values for the hopping parameters~\cite{rmp,Ando}, it
follows that the scaling relations we derive should apply
approximately over a wavevector range corresponding to
$\ln(s)_{\rm max} = \ln(q_{\rm H}/q_{\rm L}) \simeq
\ln(\gamma_{0}/\gamma_3) \simeq 2.3$.

We estimate the strength of the bare scattering amplitudes by
evaluating the 2D Coulomb scattering potential at the cut-off
wavevector $k_{\rm H}$:
\begin{equation}
\nu_0\,V_{S} \simeq \; \frac{m^*}{2\pi \hbar^2} \; \frac{2\pi
e^2}{k_{\rm H}} = \frac{\alpha_{\rm ee}}{2}
\end{equation}
where $\alpha_{\rm ee} = e^2/\hbar v_{\rm F} \simeq 2.2$ is
graphene's fine structure constant.
The value used for $V_{S}$ in the RG flows is motivated by this
estimate. The value used for $V_{D}$ is reduced by a factor of
$\exp(-k_{\rm H}d)$ compared to $V_{S}$ to account for the layer
separation $d=3.35$~\AA.

According to these estimates the bare value of $\nu_0 V_{D}$ exceeds
the stability limit of $\sim 0.6/\ln(s)_{\rm max} \sim 0.25$ by
approximately a factor of four.  The above estimates are for the
case of a graphene bilayer in vacuum.  For graphene layers on the
surface of a substrate with dielectric constant $\varepsilon$,
interactions are expected to be reduced by a factor of $\sim
(\varepsilon+1)/2$.  In the case of SiO$_2$ substrates $\varepsilon
\sim 4$ and the interaction strength exceeds the stability limit by
a much narrower margin.  We expect that additional screening effects
from graphene $\sigma$ orbitals, which are normally neglected in
continuum model calculations, will reduce interaction strengths at
wavevectors near $k_{\rm H}$ somewhat and favor stable bilayers.